\documentclass[sigconf]{acmart}
\AtBeginDocument{%
  }

\setcopyright{acmlicensed}
\copyrightyear{2026}
\acmYear{2026}
\setcopyright{cc}
\setcctype{by}
\acmConference[ICSE-SEIP '26]{2026 IEEE/ACM 48th International Conference on Software Engineering}{April 12--18, 2026}{Rio de Janeiro, Brazil}
\acmBooktitle{2026 IEEE/ACM 48th International Conference on Software Engineering (ICSE-SEIP '26), April 12--18, 2026, Rio de Janeiro, Brazil}
\acmPrice{}
\acmDOI{10.1145/3786583.3786848}
\acmISBN{979-8-4007-2426-8/2026/04}

\usepackage{graphics}
\usepackage{array}
\usepackage{longtable}
\usepackage{enumitem}

\begin{document}

\title{Beyond the Commit: Developer Perspectives on Productivity with AI Coding Assistants}

\author{Valerie Chen}
\email{valeriechen@cmu.edu}
\affiliation{%
  \institution{Carnegie Mellon University}
  \city{Pittsburgh}
  \state{PA}
  \country{USA}
}

\author{Jasmyn He}
\email{jasmyn.he@bny.com}
\affiliation{%
  \institution{BNY Mellon}
  \city{New York}
  \state{NY}
  \country{USA}
}

\author{Behnjamin Williams}
\email{behnjamin.williams@bny.com}
\affiliation{%
  \institution{BNY Mellon}
  \city{New York}
  \state{NY}
  \country{USA}
}

\author{Jason Valentino}
\email{jason.valentino@bny.com}
\affiliation{%
  \institution{BNY Mellon}
  \city{Pittsburgh}
  \state{PA}
  \country{USA}
}

\author{Ameet Talwalkar}
\email{talwalkar@cmu.edu}
\affiliation{%
  \institution{Carnegie Mellon University}
  \city{Pittsburgh}
  \state{PA}
  \country{USA}
}

\renewcommand{\shortauthors}{Chen et al.}

\begin{abstract}
Measuring developer productivity is a topic that has attracted attention from both academic research and industrial practice.
In the age of AI coding assistants, it has become even more important for both academia and industry to understand how to measure their impact on developer productivity, and to reconsider whether earlier measures and frameworks still apply.
This study analyzes the validity of different approaches to evaluating the productivity impacts of AI coding assistants by leveraging mixed-method research.
At BNY Mellon, we conduct a survey with 2989 developer responses and 11 in-depth interviews.
Our findings demonstrate that a multifaceted approach is needed to measure AI productivity impacts: survey results expose conflicting perspectives on AI tool usefulness, while interviews elicit six distinct factors that capture both short-term and long-term dimensions of productivity.
In contrast to prior work, our factors highlight the importance of long-term metrics like technical expertise and ownership of work.
We hope this work encourages future research to incorporate a broader range of human-centered factors, and supports industry in adopting more holistic approaches to evaluating developer productivity.
\end{abstract}



\received{20 February 2007}
\received[revised]{12 March 2009}
\received[accepted]{5 June 2009}

\maketitle

\section{Introduction}

AI-powered applications hold significant promise for increasing human productivity, as many models—particularly large language models (LLMs)—have demonstrated human-level capabilities in domains ranging from natural language understanding to code generation~\cite{brown2020language,chen2021_evaluating_codex}. 
One salient and increasingly common use case of AI tools in practice is that of coding assistants, with prominent examples including GitHub Copilot and Cursor. 
As these systems are integrated into real-world software development workflows, a natural question arises: how can we measure their impact on developer productivity? 
Indeed, AI coding assistants that substantially boost productivity could necessitate a rethinking of how software is developed in the age of AI~\cite{peng2023impact_copilot}.

While improving productivity is a desirable goal for AI tools, ``productivity'' has become a complex and multi-faceted construct~\cite{Forsgren2021}. 
To illustrate this challenge, consider several definitions from prior work. 
On one hand, GitHub Copilot's developer survey reports emphasize self-reported metrics such as developer satisfaction, perceived success in completing repetitive work, and ability to remain ``in the flow''~\cite{github2022copilot_survey}. 
On the other hand, field studies have considered more objective measures, such as changes in commit frequency between developers with and without access to AI coding assistants~\cite{peng2023impact_copilot,song2024generative_oss_productivity}. 
The diversity of perspectives in existing literature motivates a careful examination of the broad set of metrics that may be relevant to measuring productivity with AI tools---we refer to this as \emph{AI productivity metrics}.

While the software engineering literature offers extensive approaches to measuring developer productivity, such as the SPACE framework~\cite{Forsgren2021} and DORA metrics~\cite{forsgren2018accelerate}, comparatively little attention has been given to productivity in the context of AI coding assistants. As tools like GitHub Copilot and Cursor become increasingly embedded in everyday workflows, it is essential to examine how their properties differ from existing tools. 
AI coding assistants support a wide and expanding scope of tasks but generate outputs that are inherently non-deterministic~\citep{liang2024large}, in contrast to traditional developer aids. 
Tools such as SonarQube, a widely used static analysis system, operate on well-defined inputs and outputs, producing consistent results under the same conditions. 
Likewise, integrated development environment features such as automated code formatting (e.g., Prettier) or linting (e.g., ESLint) apply deterministic transformations and rule-based checks, guaranteeing identical outputs for identical code. 
These contrasts highlight why evaluating the productivity impacts of AI coding assistants requires new perspectives that account for their variability, unpredictability, and broader task coverage.

We approach this investigation using a mixed-methods research approach that leverages large-scale survey and semi-structured interviews with developers at BNY Mellon. 
First, we demonstrate why a holistic understanding across metrics is important by surveying nearly three thousand developers.
We deploy the Developer Experience (DX) survey framework~\cite{greiler2022actionable} across the company to collect perceptions of productivity with AI coding assistants, focusing on two core questions: developers’ satisfaction with AI tools and their perceived time savings. 
Our survey results show that, while satisfaction with coding assistants like GitHub Copilot is high, reported time savings are relatively modest. This finding underscores the need for holistic evaluation frameworks and motivates our subsequent qualitative inquiry into diverse productivity dimensions.

Next, we explore the diversity and breadth of potential productivity metrics by conducting 11 semi-structured interviews with developers of varying backgrounds and levels of seniority. 
We identify six considerations that influence productivity across development, deployment, and long-term career trajectories: \texttt{self-sufficiency}, \texttt{frustration and cognitive load}, \texttt{task completion rate}, \texttt{ease of peer review}, \texttt{technical expertise}, and \texttt{ownership of work}. 
Notably, long-term factors such as technical expertise development and sense of ownership have received little attention in prior research. These findings reinforce that productivity is inherently multi-dimensional and cannot be captured by a single metric. To contextualize these dimensions, we apply them to common use cases surfaced through the interviews, which include implementing a new feature, refactoring existing code, and generating documentation or test cases, to illustrate how AI productivity measures may have varying impacts on different use cases. 

We conclude by offering recommendations for operationalizing these factors in empirical studies and industry evaluations and providing a discussion that compares our factors to existing work on developer productivity.
We hope our work motivates and lays the the groundwork for more nuanced, context-aware productivity measurement in AI-augmented software engineering.

\section{Related Work}

We outline two areas relevant to our study: (1) productivity measures in software engineering, including traditional and contemporary frameworks, and (2) empirical work on measuring productivity with AI coding assistants.

\paragraph{Background on productivity in software engineering.}  
Productivity in software engineering has been a growing area of research, which has increasingly shown how productivity is a multi-faceted construct~\cite{Sadowski2019}. 
It has traditionally been evaluated through quantitative metrics like the ratio of output to effort, including lines of code over time, software complexity, and the presence of errors or defects~\cite{Halstead1977,McCabe1976, Ko2005,LowJeffery1990}. 
However, these measures have well-known limitations in capturing the quality and maintainability of code.  

More recent work has also emphasized subjective and cognitive dimensions, including developer perception, well-being, and satisfaction, which are now recognized as essential productivity factors~\cite{Meyer2014}. 
For example,~\citet{icse2022productivity} found that perceived productivity was the strongest predictor of team-level productivity, even more so than meetings or coordination activities. 
Similarly, organizational and individual measures of engineering success can diverge, highlighting the need to separate different layers of measurement~\cite{lee2023our}. 
Complementing these perspectives, new telemetry-driven metrics such as Diff Authoring Time (DAT) provide fine-grained signals of developer effort at scale~\cite{beller2025s}.

More recent frameworks have aimed to unify multiple facets of productivity measures.
For example, the SPACE framework distills both objective and subjective considerations across individuals, teams, and organizations~\cite{Forsgren2021}, while the DORA metrics focus on DevOps performance~\cite{ForsgrenKersten2018}. 
Despite these advances, most frameworks were developed for traditional or DevOps workflows and do not explicitly address the unique characteristics of AI coding assistants, which is the focus of our work. 

\paragraph{Measuring productivity with AI coding assistants.}  
The adoption of AI coding assistants has risen sharply in recent years, with tools such as GitHub Copilot and Cursor now integrated into major IDEs and workflows used by millions of developers~\cite{github_octoverse2023}. 
Their use spans a number of day-to-day developer tasks such as code completion, automated documentation, refactoring suggestions, and even test generation~\cite{Vaithilingam2022,ziegler2024}. 
To evaluate the impact of these tools on developer productivity, many user studies have been conducted with popular assistants like GitHub Copilot or bespoke tools powered by different LLMs. 
However, the findings are mixed.  

For example, \citet{Weisz2022} found that different users benefited to varying degrees when using AI coding assistants. 
\citet{ziegler2024} found that developers who used GitHub Copilot reported higher productivity levels. 
Comparative evaluations across multiple assistants (e.g., Copilot, TabNine, ChatGPT) further demonstrate heterogeneous outcomes in correctness, speed, and acceptance~\cite{empirical2024assistants}. 
At the same time, large-scale field studies of AI-enabled code search illustrate that adoption is uneven and effectiveness must outweigh switching costs to drive real productivity benefits~\cite{fse2024codesearch}. 
Complementary perspectives even propose physiological or cognitive-load measures, such as those gathered via wearables, as additional dimensions to capture productivity impacts of generative AI~\cite{icse2025wearables}. 
Across these studies, inconsistent metrics are employed to study productivity, leading to different downstream conclusions. 
Thus, we create a similar framework—like SPACE/DORA—but specifically tailored for AI coding assistants.

\section{Research Design}

We employed a sequential exploratory mixed-methods research design~\cite{creswell2003advanced} involving two phases. 
First, we conducted a large-scale survey of nearly three thousand developers at BNY Mellon to quantify perceptions of productivity with AI coding assistants using two different metrics. 
This quantitative phase allowed us to measure the prevalence of key developer experiences at scale. 
Next, building on these results, we conducted in-depth semi-structured interviews to qualitatively explore the nuanced productivity dimensions and contextual factors that underlie these perceptions. 
This approach combines broad quantitative insights with rich qualitative understanding to capture the complexity of AI-assisted software development to answer the following research questions:
\begin{itemize}[leftmargin=15pt]
    \item \textbf{RQ1:} To what extent do different AI productivity metrics align in their assessments?
    \item \textbf{RQ2:} What broader range of AI productivity metrics do developers consider relevant in the context of AI coding assistants?
    \item \textbf{RQ3:} How do developers view the impact of AI productivity metrics on different parts of the development process?
\end{itemize}

\subsection{Survey}\label{subsec:survey}

While prior studies have conducted surveys to understand the impact of AI coding assistants on various aspects of developer experience (e.g., ~\cite{github2022copilot_survey,mozannar2024realhumaneval,weisz2025examining}), they have not explicitly analyzed and reported how user responses across different productivity metrics, which may be used in each of these surveys, correlate with each other (\textbf{RQ1}).

\paragraph{Participants.} To recruit a varied and representative group of survey respondents, we leveraged the large pool of engineers throughout BNY Mellon who actively commit code.
The survey was completed by around 8000 total engineers throughout the company.
Engineers opted in to answering questions based on whether it was relevant to them: it is possible that an engineer who does use AI tools failed to answer the pertinent survey question.
Regarding our particular questions, we had 2989 respondents---which is a majority of engineers who have access to GitHub Copilot, a specific AI coding assistant that developers have access to within the company.
All survey responses were anonymous.

\paragraph{Protocol.} The survey was conducted using the DX platform~\cite{greiler2022actionable}, which was created by the inventors of the SPACE and DORA frameworks.
The survey contained two questions that pertain to AI coding assistants, which are the focus of our work:\footnote{The two questions presented in our study are part of a larger survey on developer experience within BNY Mellon. To protect the confidentiality of company metrics and goals, we omit the remainder of the survey questionnaire beyond these two questions.} 
\begin{enumerate}[leftmargin=15pt]
    \item \textbf{General satisfaction~\cite{StackOverflowPulse2024,github2022copilot_survey}:} Consumer Satisfaction score for GitHub Copilot. The response options are ``very dissatisfied,'', ``dissatisfied,'' ``neutral,'', ``satisfied,'' and ``very satisfied.''
    \item \textbf{Perceived time savings~\cite{JetBrainsAIAssistant2024TimeSavings}:} On average, how much time is saved with GitHub Copilot per week? The response options are ``No time savings,'' ``1-30 min per week,'' ``31-60 min per week,'' ``1-2 hours per week,'' and ``2+ hours per week.''
\end{enumerate}
\noindent These questions represent two approaches to measuring productivity tools: one looks at an objective metric (albeit estimated by the user)---similar to prior work---and the other looks at a subjective metric---which has also been considered in a separate set of prior works. We note that we considered the use of direct objective metrics, like acceptance rates or chat acceptance rates, that do not rely on self-reports, but were unable to reliably obtain these metrics across all engineers.
Participants in the survey could also elaborate on their multiple-choice responses. 

\paragraph{Analysis.} In this paper, we report primarily quantitative results from
ratings regarding both metrics.
We also include brief snippets of additional comments that were provided by survey respondents.

\subsection{Interview}

In addition to large-scale survey responses, we also want to target individual developers through semi-structured interviews to understand how they define different productivity metrics with AI coding tools. 
This helps us answer both \textbf{RQ2} and \textbf{RQ3}.

\begin{table}[t]
\centering
\caption{Interview Participants.}
\label{tab:participants}
\resizebox{\columnwidth}{!}{%
\begin{tabular}{llll}
\toprule
& \textbf{Seniority} & \textbf{Developer Role} & \textbf{Department Function} \\
\midrule
P1 & Early Career & Backend Dev & Customer Products \\
P2 & Mid Career & Full stack Dev & Platform Engineering \\
P3 & Mid Career & Full stack Dev & Data \\
P4 & Early Career & Backend Dev & Customer Products \\
P5 & Mid Career & Frontend Dev & Platform Engineering \\
P6 & Management & --- & Platform Engineering \\
P7 & Mid Career & Full stack Dev & Customer Products \\
P8 & Early Career & Backend Dev & Data\\
P9 & Mid Career & Backend Dev & Data \\
P10 & Management & --- & Platform Engineering \\
P11 & Management & --- & Platform Engineering \\
\bottomrule
\end{tabular}}
\end{table}

\begin{figure*}[t]
\centering
\includegraphics[width=\textwidth]{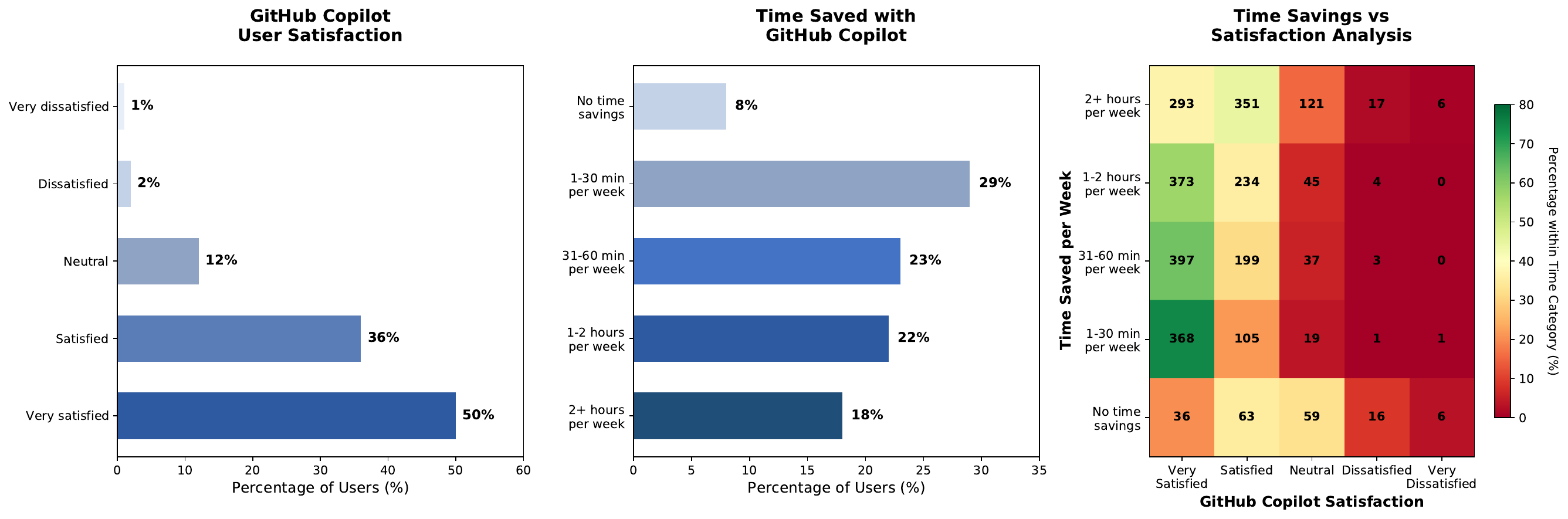}
\caption{Survey responses from 2989 engineers on productivity with GitHub Copilot. In the left and middle panels, we show two different metrics relating to general satisfaction and perceived time savings. In the right panel, we provide a fine-grained breakdown of the two metrics.} 
\label{fig:survey}
\end{figure*}

\paragraph{Participants.} We recruited developers across the company based on three different criteria to ensure a diverse set of perspectives using \emph{purposive
sampling}~\cite{campbell2020purposive}.
First, we recruited participants of different levels of seniority: early career (less than 3 years of experience), mid-career (more than 3 years of experience), and management level.
Note that even management-level technical employees still have experience with using GitHub Copilot (e.g., using them in code reviews).
We additionally considered the business sector, as the use case of AI tools may vary.
Finally, we recruited participants with varying roles in development (full stack, backend, and frontend engineers).
We obtained a list of internal developers with access to GitHub Copilot and filtered based on the list of three criteria before reaching out to them via Teams to see if they were willing to participate in the study.
We also used snowball sampling~\cite{naderifar2017snowball} to identify additional participants.
In total, we reached out to an equal number of each subgroup, but the response rate varied.
The final list of participants is in Table~\ref{tab:participants}.

\paragraph{Protocol.} We conducted 11 semi-structured interviews led by the first author. The interviews were conducted virtually via Teams, lasting between 30 and 45 minutes each.  
The interview content covered topics around the participants' engineering and AI background, use cases of GitHub Copilot, and perceptions on productivity metrics.
\begin{itemize}[leftmargin=15pt]
    \item \textbf{Engineering and AI Background.} We started each interview by asking participants to describe their current role, how long they have been at the company, and how many days a week they write code.
    We also inquired about their familiarity with GitHub Copilot and the frequency at which they use it to write code.
    For more senior-level interviewees, we asked them to describe their prior approaches to software development and how it has changed since gaining access to AI tools.
    \item \textbf{Use Cases of AI coding assistants.} We asked participants to talk through what successful and unsuccessful use cases of GitHub Copilot were throughout the software development life cycle. 
    This helps contextualize our discussion of varying productivity metrics in grounded use cases in software development.
    \item \textbf{Perceptions on productivity metrics.} We talked about how they would define productivity in the context of AI coding assistants (e.g., GitHub Copilot). 
    We also brought up sample metrics from prior work, like those used in Section~\ref{subsec:survey}, and asked them to comment on them. 
\end{itemize}
Upon completion of the interview, participants were compensated with internal company points for their time. 

\paragraph{Analysis.} For our semi-structured interview data, we recorded and transcribed each interview in addition to taking observational notes.
To answer \textbf{RQ2}, we conducted inductive open coding and applied a descriptive thematic analysis of all responses to the questions about perceptions on productivity metrics.
We organize recurring patterns in the interview data, iterating multiple times to refine the resulting set of factors impacting developer productivity~\cite{braun2006using,ahmed2025using}.
We then map the set of factors to relevant phases in the software development process to contextualize how productivity effects arise across development, deployment, and long-term maintenance.
To address \textbf{RQ3}, we also created codes corresponding to GitHub Copilot use cases mentioned by multiple participants for the case studies, focusing our discussion on the top three most common ones.
Quotations have been lightly edited for concision and grammar.

\section{Results}

\subsection{RQ1: Do AI productivity metrics align with each other?}

Figure~\ref{fig:survey} overviews the responses of 2898 engineers who answered at least one of the questions on general satisfaction and perceived time savings with GitHub Copilot.
On one hand, we find that most engineers are satisfied with their experience: 86\% are either satisfied or very satisfied (Figure~\ref{fig:survey} left).
As one survey respondent described: ``\textit{Its helping a lot in day-to-day work. It is providing to correct and help provide other solutions that reduce complexity and are more efficient}.''
On the other hand, we find that a majority of engineers (around 60\%) report less than one hour of time savings (Figure~\ref{fig:survey} middle)---this is well below the average reported number by other companies of similar sizes, according to DX. 
In the participant comments, we see that this may be due to hallucinations where ``\textit{asking it to fix it does not work and just returns the same wrong answer}'' or limitations due to certain use cases (i.e., ``\textit{does not work as well for C\#}'').

These results suggest that developers can be satisfied with AI coding assistants like GitHub Copilot without necessarily getting significant time savings.
When we correlate individual responses on the two questions ($N=2754$), we find a Pearson Correlation Score of $r=0.34$, a positive but weak correlation ($p<0.0001$).
When looking into the breakdown by category in Figure~\ref{fig:survey} (right), we can see nearly 400 developers are very satisfied with GitHub Copilot despite saving only 30 minutes per week, while there are over 100 developers who save over 2+ hours a week but are neutral or dissatisfied with the tool.
This type of cross-metric analysis demonstrates why productivity may not necessarily be captured with a single metric, motivating a follow-up interview study.


\subsection{RQ2: Which AI productivity metrics are important to developers?}

As we showed in RQ1, one or two metrics are not sufficient to capture developer perspectives.
To identify a diverse set of metrics, we interviewed 11 developers across BNY Mellon. 
We identify 6 factors to consider when measuring the impact of AI coding assistants.
We group them into the development process, the deployment process, and the long-term impact.

\subsubsection{Impact on software development}

During the development process, it is important to consider how a developer feels when they are writing code (P4) and whether the AI assistant is improving or hurting that experience.
We identify two key factors:

\paragraph{\texttt{(Factor 1) Do AI tools improve self-sufficiency?}}
A challenge that developers had before getting access to GitHub Copilot was the need to frequently context-switch when seeking help or feedback from external resources, whether it be a co-worker or Stack Overflow.
P2 shared how getting help may require \textit{``poking people individually and exhausting various resources like Google and Stack Overflow.''}
When using GitHub Copilot to write code, this is no longer the case if GitHub Copilot can easily infer what the developer wants and seamlessly provide the response or the next few lines of code.
When dealing with projects that require upgrading on the order of ``\textit{30-40 libraries,''} P5 notes that they \textit{``never visit Stack Overflow now.''}
Multiple participants noted feeling more self-sufficient with the AI assistant, likening the experience to \textit{``working with a co-worker''} (P5) or \textit{``a friend''} (P7). 
A related challenge is when working on projects that require multiple languages (e.g., both Java and Python); having GitHub Copilot reduces the amount of documentation to look at (P4). 
In summary, P10 noted that ``\textit{if the team can focus on other things, cognitive effort can be offloaded by allowing team members to focus on other work.}''

\paragraph{\texttt{(Factor 2) Do AI tools reduce the level of frustration and cognitive load?}}
Outputs from GitHub Copilot are not perfect.
As such, figuring out how to properly integrate LLM responses can add cognitive load for the developer.
P8 noted how they need to be ``\textit{really careful with copying responses, otherwise they may not realize there is another change.}''
There may also be challenges in terms of figuring out what is the best way to prompt the AI assistant for the correct outputs.
One developer (P3) reflected on the non-deterministic nature of LLMs, stating how you need to ask ``\textit{ask four or five times to get the correct answer, even if you ask the same thing.}'' 
This may be a problem, especially for junior developers, ``\textit{because they never wrote anything from scratch. Given a template, they don't know what to do with it.}''

\subsubsection{Impact on software deployment}

The impact of AI assistants can be seen not only on a developer's day-to-day workflow, but also on a team's general performance.

\paragraph{\texttt{(Factor 3) Do AI tools improve the rates of and quality of task completion?}}
Many participants noted the importance of measuring throughout (P5,9,11)---i.e., the rate at which individual or teams can complete a well-defined task, and quantifying how that is impacted by the use of AI assistants. 
This factor is most similar to metrics reported in prior work, though participants provided different suggestions to quantify this benefit.
For example, P5 said that if it previously took them ``\textit{5 days, target finish in 3 days and do 2 days of testing. If I can write it in X number of lines, can Copilot do the same thing in 1 day.}''
However, P9 noted that the number of lines is not the only benchmark of progress: it is ``\textit{not necessarily the amount of code you have written, but rather the criticality of the work and what kind of benefit the company is seeing.}'' 
Using simple metrics like the number of lines of code written does not necessarily correlate with impact.
Relatedly, P11 added that it remains a challenge within the company to quantify these metrics, rather than simply relying on developer perceptions.

\paragraph{\texttt{(Factor 4) Do AI tools positively impact the peer review process?}}
Peer review of code is critical to deploying reliable software to production.
However, the use of AI assistants in producing such code may have unforeseen impacts when it comes to the review process.
One senior manager noted that a recurring issue they observed with their junior developers when reviewing their code is that ``\textit{a problem if you optimize a piece of it}'' but it is not being optimized for the right things, leading others to spend additional time fixing these issues (P2).
Along with this same concern, P6 wondered how extensively junior engineers were testing their code, indicating that increased usage of coding assistants at an earlier stage of a career may also impact how easily others can review their code or vice versa. 
However, for senior developers, Copilot can even aid in facilitating the code review process by quickly summarizing changes (P3, 10).

\subsubsection{Longer-term Developer Impact}

Many interviewees also touched on factors that were outside of the immediate development and deployment cycle and not immediately quantifiable. 

\paragraph{\texttt{(Factor 5) Do AI tools improve technical expertise?}}
Many interviewees emphasized the importance of growth as a developer over time, rather than just an individual's output per sprint.
Interviewees touched on the importance of teaching junior developers the proper way to code---whether it is thinking about architecture patterns or infrastructure considerations---and appropriately scaffolding AI into this process remains challenging.
In fact, having access to AI too early might disrupt one's learning: ``\textit{If I have a team member who is brand new, they need to learn new technology. 
However, if the code just works, then you just accept it}'' (P10).
P7 reflected on their experience as a developer before getting access to GitHub Copilot---an experience that junior developers no longer have: ``\textit{you need to know how to analyze. We had to go through tons and tons of stack traces, which helped at a later point}.''
While P6 was concerned about similar issues, they were also optimistic about some skills that junior personnel may no longer need to learn, likening this transition phase to ``\textit{learning handwriting skills}'' which are no longer as important.
However, when used correctly, GitHub Copilot can facilitate developers in building new technical expertise.
For example, P11 described how they can easily ``\textit{look up and get examples of security topics that they are not aware of}'' with AI assistants.

\paragraph{\texttt{(Factor 6) Do AI tools impact a developer's sense of ownership of work?}}
In addition to their technical expertise, interviewees talked about the importance of feeling like they owned the code that they wrote.
In particular, developers mentioned how they valued the perceived authorship and accountability towards a piece of software, particularly how there is ``\textit{nothing like doing it yourself}'' (P1,3,5).
This can be especially helpful when there is a production issue, as developers who have a deep knowledge of the code feel like they can fix the issues faster.
For example, P9 described how ``\textit{if you wrote the majority of it, you'd know exactly where the issue is and know where to look.}''

\subsection{RQ3: How do AI productivity metrics impact the workflow?}

\begin{figure}[t]
\centering
\includegraphics[width=\columnwidth]{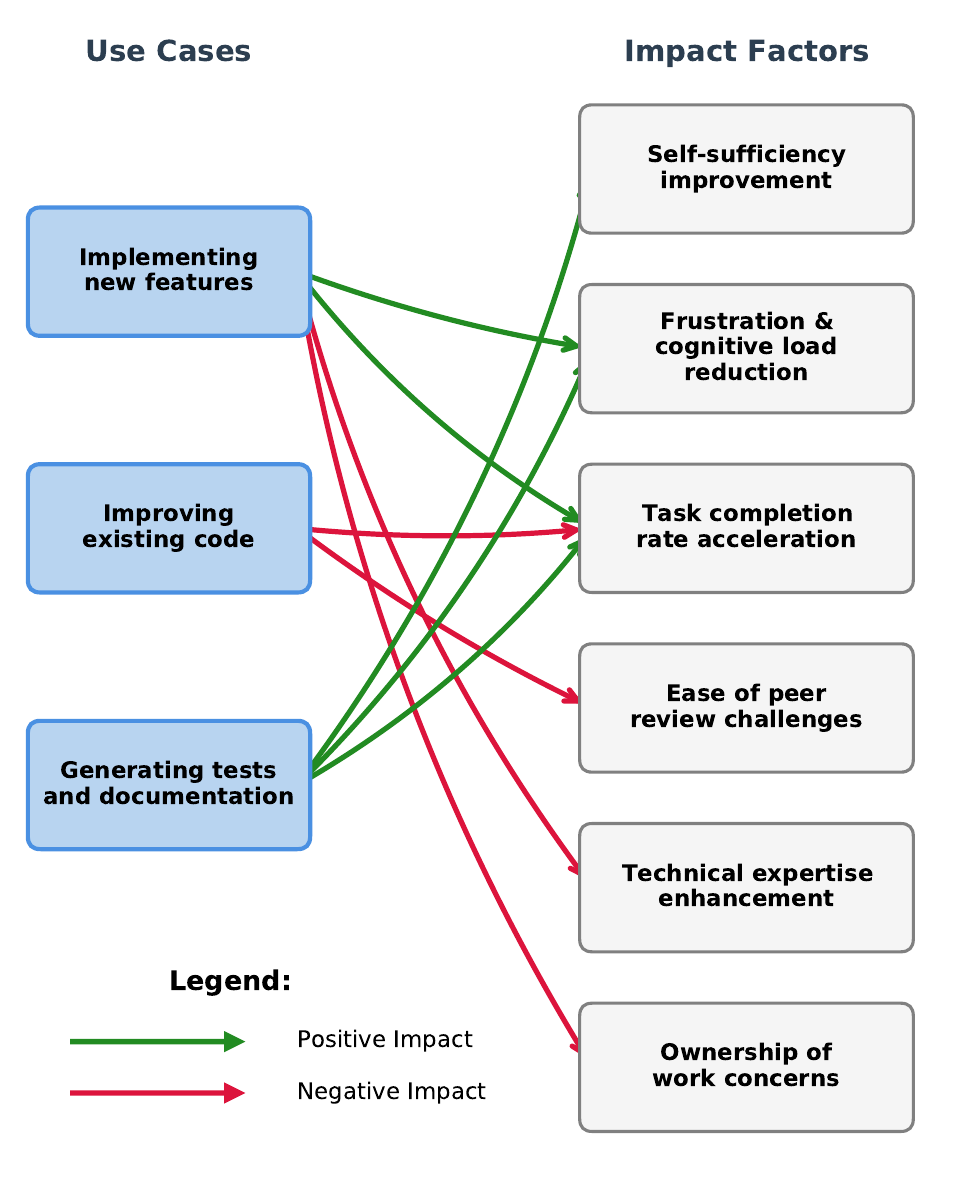}
\caption{Mapping use cases of GitHub Copilot to productivity metrics, both in terms of positive and negative impact based on interviewee sentiment.}
\label{fig:analysis}
\end{figure}

In our semi-structured interviews, developers discussed multiple use cases for GitHub Copilot in their day-to-day workflows.
We highlight three of the most mentioned use cases, discuss how GitHub Copilot impacts each use case, and analyze how the use of coding assistants impacts the proposed factors.
Overall, we find that the impact of GitHub Copilot on productivity factors varies by use cases (Figure~\ref{fig:analysis}), which highlights the importance of carefully considering the context of usage when making productivity claims.

\subsubsection{Use Case 1: Implementing new features (P1,2,3,6,8,11).}
One of the predominant use cases of GitHub Copilot is to write code to implement new features~\cite{peng2023impact_copilot}.
Interviewees described how AI can easily provide multiple implementations and \textit{``turn out far more lines of code''} than a human developer can (P6). While AI can give developers a good starting point, there are concerns about \textit{``blindly copying and pasting''} AI outputs (P1,3).
However, we see a varied impact on short versus long-term metrics in terms of implementing new features and improving existing code, particularly for developers who have not had as much experience writing code on their own.
Interviewees discussed how using GitHub Copilot may improve certain factors in the shorter term---e.g., reduce frustration and cognitive load (\texttt{Factor 2}) and development time (\texttt{Factor 3}), but expressed concern about junior developers over-relying on GitHub Copilot at the expense of long-term technical expertise (\texttt{Factor 5}) as well as ownership (\texttt{Factor 6}).

\begin{table*}[t]
\centering
\caption{We identify 6 factors impacting developer productivity with AI coding assistants that span short- to long-term dimensions. Based on participant quotes, we curate questions to facilitate future evaluations of each factor.}
\label{tab:ai_impact_factors}

\resizebox{\textwidth}{!}{%
\begin{tabular}{>{\raggedright\arraybackslash}m{5.5cm} 
                |>{\raggedright\arraybackslash}m{5.5cm} 
                |>{\raggedright\arraybackslash}m{5.5cm}}
\toprule
\textbf{Impact on Development} & \textbf{Impact on Deployment} & \textbf{Longer-term Developer Impact} \\
\midrule
\texttt{\textbf{(1) Self-sufficiency:}}
\begin{itemize}[leftmargin=15pt,nosep]
  \item What questions do developers still escalate to teammates or external resources?
  \item To what extent does AI reduce context-switching between tools (e.g., docs, Stack Overflow, Google)?
  \item How confident are developers in relying solely on AI suggestions for everyday work?
  
\end{itemize} 
& 
\texttt{\textbf{(3) Rate of task completion:}}
\begin{itemize}[leftmargin=15pt,nosep]
  \item By what measurable margin (time saved, lines of code written, bugs avoided) do tasks complete faster?
  \item  Which categories of tasks (e.g., boilerplate, bug fixing, design-heavy work) are accelerated most by AI?
  \item Do developers feel AI frees up time for more creative or higher-value work?
\end{itemize}
& 
\texttt{\textbf{(5)Technical expertise:}}
\begin{itemize}[leftmargin=15pt,nosep]
   \item Do junior engineers over-rely on AI, and how does this affect onboarding or learning?
  \item Are developers at risk of skill decay due to continued AI usage?
  \item How can management monitor and balance against potential loss of expertise?
\end{itemize} \\
\midrule
\midrule
\texttt{\textbf{(2) Frustration \& cognitive load:}}
\begin{itemize}[leftmargin=15pt]
  \item How often do developers experience frustration due to irrelevant, incorrect, or repetitive AI suggestions?
  \item How much time is spent reformulating or experimenting with prompts?
  \item How does mental fatigue compare on days with heavy AI use versus without?
\end{itemize} 
& 
\texttt{\textbf{(4) Ease of peer review:}}
\begin{itemize}[leftmargin=15pt]
  \item Does AI-generated code require more, less, or different types of review comments?
  \item Does the use of AI affect trust between reviewers and authors?
  \item Can reviewers distinguish between AI- and human-written code? Does this matter for review quality?

\end{itemize}
& 
\texttt{\textbf{(6) Ownership of work:}}
\begin{itemize}[leftmargin=15pt]
  \item Do developers maintain a deep understanding of codebases when AI contributes significantly?
  \item Does AI affect perceived authorship, accountability, and willingness to maintain code?
 \item How does reliance on AI shape long-term identity of a developer and broader team culture?

\end{itemize} \\
\bottomrule
\end{tabular}%
} 
\end{table*}

\subsubsection{Use Case 2: Improving existing code (P3,4,7,9,10).}
In addition to implementing new features, developers would also like GitHub Copilot to improve an existing code base~\citep{chen2021_evaluating_codex,shirafuji2023refactoring,cordeiro2024empirical}.
To improve existing code---e.g., refactoring a large code base or legacy applications that have been \textit{``written a bunch of ways''} (P3,4), interviewees described that the bulk of work is understanding what changes to make and not necessarily in terms of making the changes themselves, which may only be a few lines of code changes. 
P11 describes ``\textit{how 70-80\% of the time is about reading code, then only spending 10-20\% is about writing code},'' in fact not benefiting much from AI assistance if they ``\textit{already know where to put the code}.''
However, AI typically would fail without ``\textit{a lot of context}'' and ``\textit{very specific instructions}'' (P7).
Across the board, developers find it challenging to use GitHub Copilot to refactor code, particularly impacting deployment---i.e., task completion rate (\texttt{Factor 3}) and ease of peer review (\texttt{Factor 4}).

\subsubsection{Use Case 3: Generating test cases and documentation (P3,4,5,7,11).}
In addition to generating code for new features, GitHub Copilot are also used to generate tests and documentation for code~\citep{wang2024testeval,lu2021codexglue}.
Interviewees described test case generation and documentation as fairly ``\textit{boilerplate}'' tasks where developers rely on AI to automate the process (P4,7). 
As a result, P7 described how this is a task that could have taken one day to complete without AI help, but now it can be done within one hour.
Developer sentiment suggests that GitHub Copilot can be overall helpful, particularly in reducing frustration (\texttt{Factor 2}) and speeding up the rate of task completion (\texttt{Factor 3}).
However, as a word of caution in line with building appropriate technical expertise (\texttt{Factor 5}), P5 noted that ``\textit{it's still important to think through the scenarios the tests should cover}.''

\section{Discussion}

Based on our findings, we discuss how to operationalize our AI productivity factors in real-world development contexts and compare the factors we identified to other metrics that have been proposed in prior work.

\begin{table*}[t]
\caption{Comparison of our identified productivity factors with existing frameworks and prior studies.}
\centering
\small
\begin{tabular}{p{4.5cm}p{2.5cm}p{6cm}}
\toprule
\textbf{Factor} & \textbf{SPACE/DORA?} & \textbf{Prior Studies} \\
\midrule
\texttt{(1) Self-sufficiency} & No &
\begin{itemize}[leftmargin=*,itemsep=0pt]
  \item Weisz et al.~\cite{Weisz2022}: variation in how much developers relied on AI assistance
\end{itemize} \\
\midrule
\texttt{(2) Reduced cognitive load} & Yes (SPACE) 
&
\begin{itemize}[leftmargin=*,itemsep=0pt]
  \item Ziegler et al.~\cite{ziegler2024}: Copilot improved perceived efficiency and flow
  \item Mozannar et al.~\cite{mozannar2024}: assistants reduce developer effort in complex tasks
\end{itemize} \\
\midrule
\texttt{(3) Rate of task completion} & Yes (DORA) 
&
\begin{itemize}[leftmargin=*,itemsep=0pt]
  \item Imai~\cite{Imai2022}: AI tools increased lines of code but not necessarily useful output
\end{itemize} \\
\midrule
\texttt{(4) Ease of peer review} & Yes (SPACE) 
&
\begin{itemize}[leftmargin=*,itemsep=0pt]
  \item Nguyen and Nadi~\cite{NguyenNadi2022}: AI code quality issues complicating review
\end{itemize} \\
\midrule
\texttt{(5) Long-term technical expertise} & No &
N/A \\
\midrule
\texttt{(6) Long-term Ownership of work} & No &
N/A \\
\bottomrule
\end{tabular}
\label{tab:comparison}
\end{table*}

\subsection{Operationalizing AI productivity factors}

For companies interested in incorporating a more holistic evaluation of AI productivity or researchers conducting user studies evaluating LLM use in production settings, we discuss what it would take to operationalize the factors we present in this work.
We anticipate that these considerations will be increasingly important given the large number of potential AI tools for software development.

\paragraph{Periodic surveys of user experience.}
Many companies are increasingly adopting survey approaches to periodically get feedback from their developers, whether through informal forms or through more sophisticated platforms like DX.
However, even platforms like DX are still geared towards general developer experiences and there is little focus on AI coding assistants.
As we showed in the large-scale survey earlier, questions around AI coding assistants are limited to consumer satisfaction and perceived time saved.
The two questions can each be grouped under \texttt{Factor 2} and \texttt{Factor 3} respectively, though there are additional nuances that can be studied through more in-depth questionnaires.
For each factor, we highlight more granular questions that can be included in future surveys based on points raised by participants in semi-structured interviews (Table~\ref{tab:ai_impact_factors}).
Depending on the particular interest (e.g., impact on development versus deployment), the company or researcher might tailor survey questions accordingly.

\paragraph{Analyses of developer interactions with AI coding assistants.}
Beyond user surveys, objective metrics may also help measure the impact of AI coding assistants on deployment-related factors.
Commit frequency and code completion acceptance are frequently used to measure a developer's usage of AI coding assistants.
While these metrics are relevant to deployment factors, they are not able to capture concrete time spent either completing the task (\texttt{Factor 3}) or in the peer review process (\texttt{Factor 4}).
It remains that we largely rely on survey-based approaches to measure the \textit{perceived} time saved rather than the \textit{actual} time saved---prior work has shown that users themselves may not reliably report the amount of time savings.
Companies deploying these tools may benefit from scaffolding measures around \texttt{Factors 3 and 4}.

\subsection{Comparison to Prior Work}

Finally, we discuss our findings in the context of prior work on measuring developer productivity. We focus on two categories: general developer productivity and productivity with AI tools, summarized in Table~\ref{tab:comparison}.

\paragraph{Factors present in frameworks on general developer productivity.}
Existing frameworks like DORA and SPACE capture only part of the productivity factors we identify. 
DORA emphasizes throughput and stability—measures such as deployment frequency, lead time for changes, and change failure rate—which align closely with our \textit{rate of task completion} factor (\texttt{Factor 3}). 
SPACE provides broader coverage: its dimension of efficiency and flow resonates with our \textit{reduced friction / cognitive load} factor (\texttt{Factor 2}), while its emphasis on collaboration and communication overlaps with our \textit{ease of peer review} factor (\texttt{Factor 4}). 
By contrast, \textit{self-sufficiency} (\texttt{Factor 1}) is not explicitly addressed in these frameworks, likely because traditional developer tools were more well-scoped and deterministic, with minimal impacts on a developer’s independence or reliance on the tool.

\paragraph{Implications on Productivity with AI Tools.}
We find studies generally consider metrics that span \texttt{Factors 1-4} that pertain to impacts on development and deployment. 
For example, many of the aforementioned SPACE dimensions have been operationalized in GitHub's survey of Copilot users~\citep{github2022copilot_survey}.
Industry-wide studies have operationalized productivity using a range of different metrics that map onto our factors. 
For example, Ziegler et al.~\cite{ziegler2024} measured \textit{reduced friction} through self-reported efficiency and flow; 
Weisz et al.~\cite{Weisz2022} captured \textit{self-sufficiency} by examining the extent to which developers relied on AI assistance; 
Imai~\cite{Imai2022} evaluated \textit{task completion} using lines of code and functional correctness; 
and Nguyen and Nadi~\cite{NguyenNadi2022} studied \textit{peer review} challenges by analyzing code quality and reviewability of AI-generated suggestions.

\paragraph{Prior work tends to miss out on long-term factors.}
Our study highlights how long-term factors — \emph{long-term technical expertise} and \emph{ownership of work} (\texttt{Factors 5 and 6})—remain largely unaddressed in existing frameworks and surveys. 
Participants in our study voiced concerns that over-reliance on AI could erode debugging proficiency among junior engineers or weaken collective ownership of code. 
These longer-term effects are particularly salient for emerging technologies such as code agents, whose autonomy and sustained interaction with developers may amplify such dynamics. 
By extending beyond short-term outcomes to incorporate long-term risks and trade-offs, our framework complements existing frameworks and studies with qualitative insights into how AI tools reshape the developer experience over time.

\subsection{Threats to Validity}

The surveys and interviews conducted in this work were limited to one company where there was growing adoption of AI coding assistants by developers.
While the results may not generalize to all other organizations or to independent developers, we attempted to recruit a diverse set of participants that ranged in terms of sector, development expertise, and year of experience to maximize coverage.
The primary external AI coding assistant that engineers in this company have access to during the time of the study is GitHub Copilot, which is representative of many features of many current coding assistants.
However, we were not able to assess the user experience with other AI coding assistants.
While new, emerging tools may include more agentic workflows~\citep{chen2025code}, those were out of scope for this work.
Further, our findings provide a set of factors and recommendations for measuring AI productivity and how to operationalize these factors in practice.
Due to the aforementioned limitations, the factors may not be fully comprehensive.

\section{Conclusion}

Understanding productivity in the age of AI coding assistants requires rethinking established measures and frameworks. 
We leverage a mixed-methods approach to show that developer productivity using AI coding assistants cannot be captured by single metrics alone.
First, our survey results revealed high satisfaction but only modest time savings, underscoring the limits of efficiency-focused measures. 
Second, our semi-structured interviews  surfaced six factors, ranging from self-sufficiency and reduced cognitive load to long-term expertise and ownership, that capture both immediate and long-term impacts. 
By framing productivity as multi-dimensional and context-dependent, this work offers a foundation for more holistic evaluations of AI coding assistants for future research and industry deployments.

\bibliographystyle{ACM-Reference-Format}
\bibliography{sample-base}

\end{document}